\documentclass[twocolumn,showpacs,preprintnumbers,amsmath,amssymb]{revtex4}

\usepackage{graphicx}
\usepackage{dcolumn}
\usepackage{bm}

\begin{document}

\preprint{In preparation for Phys. Rev. B}

\title{
Confirmation of a one-dimensional spin-1/2 Heisenberg system with ferromagnetic first-nearest-neighbor and antiferromagnetic second-nearest-neighbor interactions in Rb${}_{2}$Cu${}_{2}$Mo${}_{3}$O${}_{12}$
}

\author{Masashi Hase}
 \email{HASE.Masashi@nims.go.jp}
\author{Haruhiko Kuroe${}^{1}$}
\author{Kiyoshi Ozawa}
\author{Osamu Suzuki}
\author{Hideaki Kitazawa}
\author{Giyuu Kido}
\author{Tomoyuki Sekine${}^{1}$}

\affiliation{%
National Institute for Materials Science (NIMS), 1-2-1 Sengen, Tsukuba 305-0047, Japan
\\${}^{1}$Department of Physics, Sophia University, 7-1 Kioi-cho, Chiyoda, Tokyo 102-8854, Japan
}%

\date{\today}

\begin{abstract}

We have investigated magnetic properties of Rb$_2$Cu$_2$Mo$_3$O$_{12}$ powder.
Temperature dependence of magnetic susceptibility and magnetic-field dependence of magnetization have shown that this cuprate is a model compound of a one-dimensional spin-1/2 Heisenberg system with ferromagnetic first-nearest-neighbor (1NN) and antiferromagnetic second-nearest-neighbor (2NN) competing interactions (competing system).
Values of the 1NN and 2NN interactions are estimated as $J_1 = -138$ K and $J_2 = 51$ K ($\alpha \equiv J_2 / J_1 = -0.37$).
This value of $\alpha$ suggests that the ground state is a spin-singlet incommensurate state.
In spite of relatively large $J_1$ and $J_2$, no magnetic phase transition appears down to 2 K, while an antiferromagnetic transition occurs in other model compounds of the competing system with ferromagnetic 1NN interaction.
For that reason, Rb$_2$Cu$_2$Mo$_3$O$_{12}$ is an ideal model compound to study properties of the incommensurate ground state that are unconfirmed experimentally.

\end{abstract}


\maketitle

\section{INTRODUCTION}

Quantum spin systems exhibit various interesting properties. They have been studied extensively.
One example of interesting spin systems is a one-dimensional spin-1/2 Heisenberg system with first- and second-nearest-neighbor interactions whose Hamiltonian is expressed as 
\begin{equation}
{\cal H} = \sum_{i=1}^{N} (J_1 S_{i} \cdot S_{i+1} + J_2 S_{i} \cdot S_{i+2}).
\end{equation}
Here $S_i$ is a spin-1/2 operator at the $i$th site, and $J_1$ or $J_2$ is a first-nearest-neighbor (1NN) or second-nearest-neighbor (2NN) exchange interaction constant.
When $J_2$ is positive (antiferromagnetic; AF), competition between the two interactions occurs irrespective of the sign of $J_1$.
Therefore, intriguing phenomena are expected to appear.
We label the spin system expressed by Eq. (1) with positive $J_2$ as a competing system in this article.

The competing system has been investigated theoretically over many years.
When both $J_1$ and $J_2$ are AF, the ground state is a spin-liquid state. A spin gap opens between the spin-singlet ground and excited states when $\alpha \equiv J_2 / J_1$ exceeds a critical value $\alpha_{\rm c}$.\cite{Haldane82}
At present, $\alpha_{\rm c}$ is evaluated as $0.24 \sim 0.30$.\cite{Tonegawa87,Affleck89,Okamoto92,Castilla95}
The exact ground state is obtained when $\alpha = 0.5$.\cite{Majumdar70,Shastry81}
The ground state is expressed by products of singlet pairs formed between nearest-neighbor spins. It has twofold degeneracy.
When $J_1$ is negative (ferromagnetic; F) and $J_2$ is AF, the ground state is the ferromagnetic state for $-0.25 < \alpha \leqslant 0$ and an incommensurate state with $S_{\rm tot} = S_{\rm tot}^{z} = 0$ for $\alpha < -0.25$.\cite{Tonegawa89}
Here $S_{\rm tot}$ and $S_{\rm tot}^{z}$ are the total spin and its $z$-component.
When $\alpha < -0.25$, it has been suggested that the gap is strongly reduced to the extent that the gap is too small for observation by any numerical method.\cite{Itoi01}
The exact ground state is obtained when $\alpha = -0.25$,  according to Hamada et al.\cite{Hamada88}
A state with $S_{\rm tot} = S_{\rm tot}^{z} = 0$ and $N+1$ states with $S_{\rm tot} = N/2$ and $S_{\rm tot}^{z} = 0, \pm 1, \pm 2, ... \pm N/2$ (ferromagnetic states) are degenerate in energy and become the ground state.
The state with $S_{\rm tot} = S_{\rm tot}^{z} = 0$ is expressed by a linear combination of states of products of all singlet pairs which are distributed uniformly on all lattice sites.
Hamada et al. called this state the uniformly distributed resonating valence bond (UDRVB) state.
The spin-singlet ground state at $\alpha < -0.25$ approaches the UDRVB state in the limit of $\alpha \to -0.25$.\cite{Tonegawa89}
Sun {\it et al.} have conjectured the existence of a new phase in the region of $-(\pi -1)/2(\pi +1) < \alpha < -0.25$ where the ground state is incommensurate and has a nonzero total spin magnitude (partially ferromagnetic polarized state).\cite{Sun02}

The first realization of the competing system is the spin-Peierls cuprate CuGeO$_3$.
The first paper reporting the appearance of the spin-Peierls transition,\cite{Hase93a} indicated that magnetic susceptibility of CuGeO$_3$ does not agree with the calculated susceptibility of a one-dimensional spin-1/2 Heisenberg antiferromagnetic system.
At first, this discrepancy had not been solved by experimental work on magnetic properties of pure and doped CuGeO$_3$.\cite{Hase93b,Hase93c,Hase95,Hase96}
Afterward, the possibility of existence of antiferromagnetic $J_2$ in addition to antiferromagnetic $J_1$ was suggested.\cite{Castilla95,Lorenzo94}
The calculated susceptibility of the competing system with antiferromagnetic $J_1$ and $J_2$ was sufficient to explain the experimental one.\cite{Castilla95,Riera95}
Until now, several model compounds of the competing system have been found.\cite{Kikuchi00,Maeshima03,Hase03,Masuda04,Kamieniarz02,Mizuno98,Matsuda95}
They are summarized in Table I.
Nevertheless, in compounds with antiferromagnetic $J_1$, the spin gap expected in the case that $\alpha > \alpha_c$ has not been confirmed experimentally.
In compounds with ferromagnetic $J_1$ already reported, values of $\alpha$ imply that the ground state is incommensurate.
These compounds are not suitable for study of the incommensurate ground state because antiferromagnetic long-range order appears at low temperature.
Therefore, discovery of further model compounds is desired because it expands experimental studies on the competing system and stimulates further theoretical interest.
A typical example is development of understanding of quantum spin systems after the observation of the spin-Peierls transition in CuGeO$_3$.\cite{Hase93a}

We have investigated several cuprates having spiral or zigzag chains of Cu$^{2+}$ ions ($S = 1/2$) in order to find model compounds including the competing system.
Recently, we reported Cu${}_{6}$Ge${}_{6}$O${}_{18}$-$x$H${}_{2}$O ($x = 0 \sim 6$) as one model compound.\cite{Hase03}
This cuprate has spiral chains of Cu$^{2+}$ ions. The chains are coupled to one another by an interchain exchange interaction.
Magnetic susceptibility of Cu${}_{6}$Ge${}_{6}$O${}_{18}$-$x$H${}_{2}$O above AF transition temperature ($T_{\rm N}$) was consistent with susceptibility obtained from the competing system with antiferromagnetic $J_1$, but an AF transition occurred at low temperature.
In addition, we obtained an experimental result suggesting the existence of a spin gap, but we were unable to prove it because of an AF transition.
In this article, we will show that Rb$_2$Cu$_2$Mo$_3$O$_{12}$, which has zigzag chains of Cu$^{2+}$ ions, is a model compound including the competing system with ferromagnetic $J_1$.

\begin{table*}
\caption{\label{table1}
Model compounds including the competing system.
$J_1$ or $J_2$ is a first- or second-nearest-neighbor interaction constant; $\alpha$ is defined as $J_2 / J_1$.
$T_{\rm N}$ indicates the AF transition temperature.
}
\begin{ruledtabular}
\begin{tabular}{ccccc}
& $J_1$ (K) & $J_2$ (K) & $\alpha$ & $T_{\rm N}$ (K)\\
\hline
CuGeO$_3$\footnotemark[1] & $150 \sim 160$ & $36 \sim 57.6$ & $0.24 \sim 0.36$ & SP\\
Cu(ampy)Br$_2$\footnotemark[2] & 17 & 3.4 & 0.2 & \\
(N$_2$H$_5$)CuCl$_3$\footnotemark[3] & 4.1 & 16.3 & 4 & 1.55\\
Cu${}_{6}$Ge${}_{6}$O${}_{18}$-6H${}_{2}$O\footnotemark[4] & 222 & 60 & 0.27 & 38.5\\
Cu${}_{6}$Ge${}_{6}$O${}_{18}$-0H${}_{2}$O\footnotemark[4] & 451 & 131 & 0.29 & 73.5\\
Li$_{1.16}$Cu$_{1.84}$O$_{2.01}$\footnotemark[5] & 67 & 19 & 0.29 & 22.3\\
Pb[Cu(SO$_4$)(OH)$_2$]\footnotemark[6] & -30 & 15 & -0.5 & 2\\
La$_6$Ca$_8$Cu$_{24}$O$_{41}$\footnotemark[7] & -215 & 78 & -0.36 & 12.2\footnotemark[8]\\
Li$_2$CuO$_2$\footnotemark[7] & -100 & 62 & -0.62 & 8.3\footnotemark[9]\\
Ca$_2$Y$_2$Cu$_5$O$_{10}$\footnotemark[7] & -25 & 55 & -2.2 & 29.5\footnotemark[10]\\
Rb$_2$Cu$_2$Mo$_3$O$_{12}$\footnotemark[11] & -138 & 51 & -0.37 & \\
SrCuO$_2$\footnotemark[12] & & 1800 & $10 \sim 1000$ & 2\\
\end{tabular}
\end{ruledtabular}
\footnotetext[1]{Ref.~\onlinecite{Castilla95,Riera95}. 
SP indicates occurrence of the spin-Peierls transition.}
\footnotetext[2]{Ref.~\onlinecite{Kikuchi00}. Cu[2-(2-aminomethyl)pyridine]Br$_2$ is abbreviated to Cu(ampy)Br$_2$. No magnetic phase transition is seen down to 1.6 K.}
\footnotetext[3]{Ref.~\onlinecite{Maeshima03}.}
\footnotetext[4]{Ref.~\onlinecite{Hase03}.}
\footnotetext[5]{Ref.~\onlinecite{Masuda04}. The magnetic structure at low temperature is helimagnetic.}
\footnotetext[6]{Ref.~\onlinecite{Kamieniarz02}.}
\footnotetext[7]{Ref.~\onlinecite{Mizuno98}.}
\footnotetext[8]{Ref.~\onlinecite{Matsuda96}.}
\footnotetext[9]{Ref.~\onlinecite{Sapina90}.}
\footnotetext[10]{Ref.~\onlinecite{Matsuda98}.}
\footnotetext[11]{This work. No magnetic phase transition is seen down to 2 K.}
\footnotetext[12]{Ref.~\onlinecite{Matsuda95}. The value of $\alpha$ in this table is the estimated magnitude of $\alpha$ because the sign of $J_1$ is not determined.}
\end{table*}

\section{Crystal structure and spin system of Rb$_2$Cu$_2$Mo$_3$O$_{12}$}

Solodovnikov and Solodovnikova first synthesized Rb$_2$Cu$_2$Mo$_3$O$_{12}$ and determined its crystal structure.\cite{Solodovnikov97}
The space group is monoclinic C2/c (No. 15).
Lattice parameters are $a = 27.698$ \AA, $b = 5.1018$ \AA, $c = 19.292$ \AA, and $\beta = 107.256^{\circ}$ with $Z = 8$ Rb$_2$Cu$_2$Mo$_3$O$_{12}$ formula units per unit cell at room temperature.
Localized spins exist only on Cu${}^{2+}$ ions ($S = 1/2$). Their positions are shown schematically in Fig. 1(a).
There are two crystallographic Cu sites.
Slightly distorted chains formed by edge-shared CuO$_6$ octahedra parallel to the $b$ axis correspond to $S = 1/2$ zigzag chains.
The 1NN Cu-Cu bond in the chains (bold bars in Fig. 1) has a slight alternation: a Cu-Cu distance is 3.08 \AA \ and Cu-O-Cu angles are 90.1 and 102.0$^{\circ}$ in one bond; and the distance is 3.09 \AA \ and the angles are 92.0 and 101.2$^{\circ}$ in the other bond.
We assume that the exchange interactions in these bonds $J_1$ are the same because the difference in the distances and angles between the two bonds is small.
As shown later, experimental results and calculated ones based on this assumption are not mutually contradictory.
The sign of $J_1$ cannot be determined from the crystal structure because both cases are allowed in such Cu-O-Cu angles.
Because the Cu-O-Cu angle is in the vicinity of 90$^{\circ}$, the exchange interaction in the 2NN Cu-Cu bonds $J_2$ (thin bars in Fig. 1; 5.10 \AA) in the chains is expected to exist through Cu-O-O-Cu paths like the spin-Peierls compound CuGeO$_3$.
According to theoretical results of Mizuno et al.,\cite{Mizuno98} the sign of $J_2$ is presumed to be AF.
On the other hand, Cu-Cu distances in the other bonds except for the 1NN bond are larger than 4.90 \AA. 
The Cu-O-Cu or Cu-O-O-Cu paths bringing magnetic interactions with magnitude comparable to $J_1$ or $J_2$ are not expected in these bonds.
Consequently, Rb$_2$Cu$_2$Mo$_3$O$_{12}$ is probably a model compound including the competing system that is represented schematically in Fig. 1(b).

\section{Methods of Experiments and Calculation}

Crystalline powder of Rb$_2$Cu$_2$Mo$_3$O$_{12}$ was synthesized by solid-state reaction method.
A stoichiometric mixture of Rb$_2$CO$_3$ (2N purity), CuO (4N purity), and MoO$_3$ (5N purity) was sintered at 733 K for 260 h in air with intermittent regrinding.
We measured X-ray diffraction patterns at room temperature.
The main phase is Rb$_2$Cu$_2$Mo$_3$O$_{12}$, but a small amount of Rb$_2$Mo$_3$O$_{10}$ (nonmagnetic) was detected.
Therefore, a small amount of CuO (antiferromagnet) probably exists, but peaks of CuO are not observed as independent peaks.
Notwithstanding, effects of the impurities are negligible because the magnetic susceptibility of Rb$_2$Cu$_2$Mo$_3$O$_{12}$ is much larger than those of the impurities.

Dependence of magnetic susceptibility [$\chi (T)$] on temperature ($T$) was measured using a superconducting quantum interference device magnetometer (MPMSX L; Quantum Design). 
Dependence of magnetization [$M(H)$] on the magnetic field ($H$) was measured using an extraction-type magnetometer in $H$ up to 30 T induced by a hybrid magnet at the High Magnetic Field Center, NIMS.
Electron spin resonance (ESR) measurements were performed using an X-band spectrometer (JES-RE3X; JEOL) at room temperature with a typical resonance frequency of 9.46 GHz. 
The powder-averaged gyromagnetic ratio of Cu$^{2+}$ ($g$) was 2.03.

We calculated all energy levels in the competing system with $10 \leqslant N \leqslant 16$ under the periodic boundary condition by means of exact diagonalization. We then calculated dependence of magnetic susceptibility on temperature and dependence of magnetization on the magnetic field.
Details of the calculation method are described in Ref.~[\onlinecite{Kuroe97}].

\section{Results and discussion}

The solid curve in Fig. 2 represents magnetic susceptibility $\chi (T)$ of Rb$_2$Cu$_2$Mo$_3$O$_{12}$ powder measured in $H = 0.1$ T.
The susceptibility is defined as $M(H)/H$.
As will be shown later in Fig. 4, $M(H)$ is linearly proportional to $H$ below 1 T.
We can see a broad maximum around $T_{\rm max} = 14.3$ K in the experimental $\chi (T)$.
The susceptibility decreases with a decrease in $T$ at low temperature, but the susceptibility does not appear to reach 0 at 0 K.
No magnetic phase transition is detected to 2 K.
The broad maximum does not mean occurrence of an AF transition because $\chi (T)$ at 2 K is smaller than half of $\chi (T)$ at $T_{\rm max}$ ($\chi_{\rm max}$).
In an AF transition, on the other hand, $\chi (T)$ at sufficiently small $T$ is about two thirds of $\chi (T)$ at AF transition temperature $T_{\rm N}$ in powder samples.
Therefore, the broad maximum suggests existence of a low-dimensional AF spin system. 
The three dashed curves show calculated $\chi (T)$ of the competing system.
Parameters are $J_1 = 22.3$ K and $\alpha = 0$ for curve 1 (the Bonner-Fisher curve), and $J_1 = 29.5$ K and $\alpha = 0.24$ for curve 2.
For curves 1 and 2, the values of $J_1$ are determined such that $T_{\rm max}$ of the experimental $\chi (T)$ agrees with that of the calculated $\chi (T)$.
Curve 3 is explained later.
In all three calculated curves, the $g$ value is 2.03, and the value of the other parts ($\chi_{\rm const}$) of susceptibility, except for spin susceptibility, is assumed to be $1.5 \times 10^{-4}$ (emu/Cu mol).
Curves 1 and 2 do not agree with the experimental $\chi (T)$.
Because temperature dependence of calculated $\chi (T)$ becomes weak with an increase in $\alpha$ for $\alpha < 1$, the competing system with $\alpha < 1$ cannot explain the experimental $\chi (T)$.
Similarly, the competing system with $\alpha > 1$ does not reproduce $\chi (T)$ of Rb$_2$Cu$_2$Mo$_3$O$_{12}$ because calculated $\chi (T)$ decreases by introduction of $J_1$ to two decoupled AF chains formed by $J_2$.\cite{Maeshima03}
The fact that the calculated $\chi (T)$ of the competing system with antiferromagnetic $J_1$ are smaller than the experimental $\chi (T)$ suggests the existence of ferromagnetic interaction.
In addition, as mentioned above, $J_2$ is considered to be AF.
Consequently, a remaining possibility is the case that $J_1$ is F and $J_2$ is AF.

In order to confirm whether the experimental $\chi (T)$ can be explained by the competing system with ferromagnetic $J_1$ and antiferromagnetic $J_2$, we calculated susceptibility. 
Figure 3 shows examples where $\alpha = -0.37$ and $N = 12 \sim 16$.
As described later, the calculated $\chi (T)$ with $\alpha = -0.37$ is consistent with the experimental $\chi (T)$.
When $T/|J_1| \ge 0.1$, susceptibilities of $N = 12 \sim 16$ agree with one another, indicating the susceptibility of $N \rightarrow \infty$.
On the other hand, susceptibility at $T/|J_1| < 0.1$ does not converge.
We performed finite-size scaling to estimate the susceptibility of $N \rightarrow \infty$, but failed to estimate it.
The ground state of the competing system at $\alpha < -0.25$ is incommensurate. For that reason, we infer that $N = 16$ is insufficient to obtain susceptibility at low temperature.
Therefore, we compare the experimental susceptibility with the calculated one at $T/|J_1| \ge 0.1$.
We could not determine the value of $T_{\rm max}$ in our calculation. 
However, a broad maximum in susceptibility of the competing system exists, as indicated by a broad maximum that is visible in the susceptibility that was calculated by another group.\cite{Thanos99}
Therefore, existence of the broad maximum in $\chi (T)$ of Rb$_2$Cu$_2$Mo$_3$O$_{12}$ is consistent with the calculated result in the competing system.

We compared the experimental $\chi (T)$ with the calculated $\chi (T)$, but we were unable to determine values of $J_1$ and $\alpha$ uniquely in susceptibility.
For that reason, we evaluated those values through comparison between experimental and calculated magnetization.
Figure 4 shows magnetization at 2.6 K.
The experimental $M(H)$ indicated by the dashed curve starts to be saturated around 14 T, but is not saturated perfectly until 30 T.
Dotted and solid curves represent calculated $M(H)$ of $N = 12$ or 16 when $J_1 = -138$ K and $\alpha = -0.37$.
In contrast to susceptibility at low temperature, convergence of the calculated magnetization is sufficient at $N \geqslant 12$.
Therefore, we considered a calculated curve with $N = 16$ agrees with magnetization of the infinite chain.
Consistency between the experimental and calculated $M(H)$ is well below 12 T.
On the other hand, above 12 T, deviation appears between the experimental and calculated $M(H)$.
This deviation cannot be explained by $M_{\rm const} \equiv \chi_{\rm const} H$ with $\chi_{\rm const} = 1.5 \times 10^{-4}$ (emu/Cu mol) indicated by the dash-dotted curve in Fig. 4 because the slope of the experimental $M(H)$ above 14 T in the unit of emu/Cu mol is ten times larger than $\chi_{\rm const}$.
This deviation is probably caused by other interactions aside from $J_1$ and $J_2$.
We also calculated magnetization with $J_1 = 22.3$ K and $\alpha = 0$ or $J_1 = 29.5$ K and $\alpha = 0.24$ (not shown).
Calculated susceptibility with these values was shown in Fig. 2 and did not agree with the experimental susceptibility.
The calculated magnetization is not saturated even at 30 T and is much different from the experimental magnetization.

We investigated whether the competing system with $J_1 = -138$ K and $\alpha = -0.37$ could also explain the experimental $\chi (T)$.
Calculated $\chi (T)$ with these values is shown in Fig. 2 by the dashed curve 3. It agrees well with the experimental $\chi (T)$ in the compared region.
As a result, susceptibility and magnetization suggest that Rb$_2$Cu$_2$Mo$_3$O$_{12}$ is a model compound of the competing system with ferromagnetic 1NN and antiferromagnetic 2NN interactions.
From the value of $\alpha$, the ground state of the spin system in Rb$_2$Cu$_2$Mo$_3$O$_{12}$ is an incommensurate state with $S_{\rm tot} = S_{\rm tot}^{z} = 0$. There is a strongly reduced spin gap that is too small to be observed using any numerical method.
The small susceptibility at low temperature in comparison with $\chi_{\rm max}$ may reflect the ground state and very small spin gap.
Discrepancy between the experimental and calculated $\chi (T)$ may appear at lower temperature, which is probably attributable to other interactions aside from $J_1$ and $J_2$.

\section{Summary}

We measured temperature dependence of magnetic susceptibility and magnetic-field dependence of magnetization of Rb$_2$Cu$_2$Mo$_3$O$_{12}$ powder.
Comparison of experimental and calculated results revealed that this cuprate is a model compound of a one-dimensional spin-1/2 Heisenberg system with ferromagnetic first-nearest-neighbor and antiferromagnetic second-nearest-neighbor competing interactions (competing system).
The values of the exchange interactions were evaluated as $J_1 = -138$ K and $J_2 = 51$ K ($\alpha \equiv J_2 /J_1 = -0.37$).
The value of $\alpha$ indicates that the ground state is a spin-singlet incommensurate state.
No magnetically ordered phase was observed down to 2 K, which is much smaller than the values of $J_1$ and $J_2$.
In contrast, other model compounds of the competing system with ferromagnetic $J_1$ exhibit an AF transition.
Therefore, Rb$_2$Cu$_2$Mo$_3$O$_{12}$ is a most suitable material to investigate the incommensurate ground state that is expected theoretically, but unconfirmed experimentally in the competing system.
Future studies must address internal magnetic fields at low temperature by NMR or $\mu$SR measurements and low-lying excited states by neutron-scattering measurements.

\begin{acknowledgments}

We are grateful 
to S. Komatsu 
for synthesis of samples and x-ray diffraction measurements, 
to A. P. Tsai and T. J. Sato 
for susceptibility measurements, 
and 
to H. Yamazaki 
for ESR measurements.
This work was supported by a grant for basic research from Materials Laboratory, National Institute for Materials Science (NIMS).

\end{acknowledgments}

\newpage 

\begin{references}

\bibitem{Haldane82}
F. D. M. Haldane, Phys. Rev. B {\bf 25}, 4925 (1982); ibid {\bf 26}, 5257 (1982).

\bibitem{Tonegawa87}
T. Tonegawa and I. Harada, J. Phys. Soc. Jpn. {\bf 56}, 2153 (1987).

\bibitem{Affleck89}
I. Affleck, D. Gepner, H. J. Schultz, and T. Ziman, J. Phys. A {\bf 22}, 511 (1989).

\bibitem{Okamoto92}
K. Okamoto and K. Nomura, Phys. Lett. A {\bf 169}, 433 (1992).

\bibitem{Castilla95}
G. Castilla, S. Chakravarty, and V. J. Emery, Phys. Rev. Lett. {\bf 75}, 1823 (1995).

\bibitem{Majumdar70}
C. K. Majumdar, J. Phys. C {\bf 3}, 911 (1970); C. K. Majumdar and D. K. Ghosh, J. Math. Phys. {\bf 10}, 1388 (1969); ibid. {\bf 10}, 1399 (1969).

\bibitem{Shastry81}
B. S. Shastry and B. Sutherland, Phys. Rev. Lett. {\bf 47}, 964 (1981).

\bibitem{Tonegawa89}
T. Tonegawa and I. Harada, J. Phys. Soc. Jpn. {\bf 58}, 2902 (1989).

\bibitem{Itoi01}
C. Itoi and S. Qin, Phys. Rev. B {\bf 63}, 224423 (2001).

\bibitem{Hamada88}
T. Hamada, J. Kane, S. Nakagawa, and Y. Natsume, J. Phys. Soc. Jpn. {\bf 57}, 1891 (1988).

\bibitem{Sun02}
L. Sun, J. Dai, S. Qin, and J. Zhang, Phys. Lett. A {\bf 294}, 239 (2002).

\bibitem{Hase93a}
M. Hase, I. Terasaki, and K. Uchinokura, Phys. Rev. Lett. {\bf 70}, 3651 (1993).

\bibitem{Hase93b}
M. Hase, I. Terasaki, Y. Sasago, K. Uchinokura, and H. Obara, Phys. Rev. Lett. {\bf 71}, 4059 (1993).

\bibitem{Hase93c}
M. Hase, I. Terasaki, K. Uchinokura, M. Tokunaga, N. Miura, and H. Obara, Phys. Rev. B {\bf 48}, 9616 (1993).

\bibitem{Hase95}
M. Hase, N. Koide, K. Manabe, Y. Sasago, K. Uchinokura, and A. Sawa, Physica B {\bf 215}, 164 (1995).

\bibitem{Hase96}
M. Hase, K. Uchinokura, R. J. Birgeneau, K. Hirota, and G. Shirane, J. Phys. Soc. Jpn. {\bf 65}, 1392 (1996).

\bibitem{Lorenzo94}
J. E. Lorenzo, K. Hirota, G. Shirane, J. M. Tranquada, M. Hase, K. Uchinokura, H. Kojima, I. Tanaka, and Y. Shibuya, Phys. Rev. B {\bf 50}, 1278 (1994).

\bibitem{Riera95}
J. Riera and A. Dobry, Phys. Rev. B {\bf 51}, 16098 (1995).

\bibitem{Kikuchi00}
H. Kikuchi, H. Nagasawa, Y. Ajiro, T. Asano, and T. Goto, Physica B {\bf 284-288}, 1631 (2000).

\bibitem{Maeshima03}
N. Maeshima, M. Hagiwara, Y. Narumi, K. Kindo, T. C. Kobayashi, and K. Okunishi, J. Phys. Condens. Metter {\bf 15}, 3607 (2003).

\bibitem{Hase03}
M. Hase, K. Ozawa, and N. Shinya, Phys. Rev. B {\bf 68}, 214421 (2003); M. Hase, K. Ozawa, and N. Shinya, to be published in J. Magn. Magn. Mater. (2004).

\bibitem{Masuda04}
T. Masuda, A. Zheludev, A. Bush, M. Markina, and A. Vasiliev, to be published in Phys. Rev. Lett.

\bibitem{Kamieniarz02}
G. Kamieniarz, M. Bielinski, G. Szukowski, R. Szymczak, S. Dyeyev, and J.-P. Renard, Computer Phys. Commun. {\bf 147}, 716 (2002).

\bibitem{Mizuno98}
Y. Mizuno, T. Tohyama, S. Maekawa, T. Osafune, N. Motoyama, H. Eisaki, and S. Uchida, Phys. Rev. B {\bf 57}, 5326 (1998).

\bibitem{Matsuda95}
M. Matsuda and K. Katsumata, J. Magn. Magn. Mater. {\bf 140-145}, 1671 (1995); M. Matsuda, K. Katsumata, K. M. Kojima, M. Larkin, G. M. Luke, J. Merrin, B. Nachumi, Y. J. Uemura, H. Eisaki, N. Motoyama, S. Uchida, and G. Shirane, Phys. Rev. B {\bf 55}, R11953 (1997).

\bibitem{Matsuda96}
M. Matsuda, K. Katsumata, T. Yokoo, S. M. Shapiro, and G. Shirane, Phys. Rev. B {\bf 54}, R15626 (1996).

\bibitem{Sapina90}
F. Sapina, J. Rodriguez-Carvajal, M. J. Sanchis, R. Ibanez, A. Beltran, and D. Beltran, Solid State Commun. {\bf 74}, 779 (1990).

\bibitem{Matsuda98}
M. Matsuda and K. Katsumata, J. Magn. Magn. Mater. {\bf 177}, 683 (1998).

\bibitem{Solodovnikov97}
S. F. Solodovnikov and Z. A. Solodovnikova, Zh. Strukt. Khim. {\bf 38}, 914 (1997) [J. Struct. Chem. {\bf 38}, 765 (1997)].

\bibitem{Kuroe97}
H. Kuroe, J. Sasaki, T. Sekine, N. Koide, Y. Sasago, K. Uchinokura, and M. Hase, Phys. Rev. B {\bf 55}, 409 (1997).

\bibitem{Thanos99}
S. Thanos and P. Moustanis, Physica A {\bf 271}, 418 (1999).

\end{references}

\begin{figure} 
\caption{
(a) 
Schematic drawing of Cu$^{2+}$-ion positions ($\bigcirc$) in Rb$_2$Cu$_2$Mo$_3$O$_{12}$.
Bold and thin bars indicate the 1NN and 2NN Cu-Cu bonds in chains.
(b) 
An illustration of the spin system in Rb$_2$Cu$_2$Mo$_3$O$_{12}$.
$J_1$ and $J_2$ are exchange interaction constants in the 1NN and 2NN Cu-Cu bonds.
}
\label{Fig. 1}
\end{figure}

\begin{figure} 
\caption{
Temperature dependence of susceptibility of Rb$_2$Cu$_2$Mo$_3$O$_{12}$ powder (solid curve) and calculated values obtained from the competing system (dashed curves). 
The parameters are $J_1 = 22.3$ K and $\alpha = 0$ (Bonner-Fisher curve), $J_1 = 29.5$ K and $\alpha = 0.24$, and $J_1 = -138$ K and $\alpha = -0.37$ in curves 1, 2, and 3, respectively. 
The powder-averaged g value determined by ESR measurements is 2.03 at room temperature.
In calculated curves, the value of the other parts ($\chi_{\rm const}$) of susceptibility aside from spin susceptibility is assumed to be $1.5 \times 10^{-4}$ (emu/Cu mol).
The inset shows susceptibility below 50 K.
}
\label{Fig. 2}
\end{figure}

\begin{figure} 
\caption{
Calculated susceptibilities of the competing system with $\alpha = -0.37$ at $T/|J_1| \leqslant 1$ or $\leqslant 0.2$ in (a) or (b).
}
\label{Fig. 3}
\end{figure}

\begin{figure} 
\caption{
Magnetic-field dependence of magnetization of Rb$_2$Cu$_2$Mo$_3$O$_{12}$ powder (dashed curve) and calculated ones obtained from the competing system with $N = 12$ or $N = 16$ (dotted or solid curve) at 2.6 K.
The parameters are $J_1 = -138$ K and $\alpha = -0.37$ in the calculated curves. 
The powder-averaged g value determined by ESR measurements is 2.03 at room temperature.
The dash-dotted curve corresponds to $M_{\rm const} \equiv \chi_{\rm const} H$ with $\chi_{\rm const} = 1.5 \times 10^{-4}$ (emu/Cu mol).
}
\label{Fig. 4}
\end{figure}

\end{document}